\documentclass[singlecolumn,amsmath,amssymb]{revtex4}

\usepackage{graphicx}
\usepackage{bm}

\begin{document}

\title{Spin fluctuations from Bogoliubov Fermi surfaces in the superconducting state of S-substituted FeSe}
\author{Zhongyu Yu$^1$\footnote{ These authors contributed equally to this work.}, Koya Nakamura$^1$\footnotemark[1], Kazuya Inomata$^1$, Xiaoling Shen$^{2, 3}$, Taketora Mikuri$^3$, Kohei Matsuura$^4$\footnote{present address: Department of Applied Physics, University of Tokyo, 7-3-1 Hongo, Bunkyou-ku, Tokyo 113-8656, Japan} Yuta Mizukami$^4$\footnote{present address: Graduate School of Science, Department of Physics, Tohoku University, Sendai, Miyagi 980-857, Japan}, Shigeru Kasahara$^5$\footnote{present address: Research Institute for Interdisciplinary Science, Okayama University, Okayama 700-8530, Japan}, Yuji Matsuda,$^5$ Takasada Shibauchi,$^4$ Yoshiya Uwatoko,$^3$ and Naoki Fujiwara$^1$\footnote{Corresponding author: fujiwara.naoki.7e@kyoto-u.ac.jp}}
\affiliation{$^1$Graduate School of Human and Environmental Studies, Kyoto University, Yoshida-Nihonmatsu-cho, Sakyo-ku, Kyoto 606-8501, Japan\\
$^2$Department of Physics, Yokohama National University, 79-1 Tokiwadai, Hodogaya-ku, Yokohama, Kanagawa 240-8501, Japan\ \\
$^3$Institute for Solid State Physics, University of Tokyo, 5-1-5 Kashiwanoha, Kashiwa, Chiba 277-8581, Japan\ \\
$^4$Department of Advanced Materials Science, University of Tokyo, 5-1-5 Kashiwanoha, Kashiwa, Chiba 277-8561, Japan\\
$^5$Division of Physics, Kyoto University, Kitashirakawa Oiwake-cho, Sakyo-ku, Kyoto 606-8502, Japan}


\begin{abstract}
\bf The study of the iron-based superconductor, FeSe, has resulted in various topics, such as the interplay among superconductivity, nematicity, and magnetism, Bardeen-Cooper-Schrieffer Bose-Einstein-condensation (BCS-BEC) crossover, and Fulde-Ferrell-Larkin-Ovchinnikov (FFLO) superconductivity. Recently, topologically protected nodal Fermi surfaces, referred to as Bogoliubov Fermi surfaces (BFSs), have garnered much attention. A theoretical model for the S-substituted FeSe system demonstrated that BFSs can manifest under the conditions of  spin-orbit coupling, multi-band systems, and superconductivity with time-reversal symmetry breaking. Here we report the observation of spin fluctuations originating from BFSs in the superconducting (SC) state via $^{77}$Se-nuclear magnetic resonance measurements to 100 mK. In a heavily S-substituted FeSe, we found an anomalous enhancement of low-energy spin fluctuations deep in the SC state, which cannot be explained by an impurity effect. Such unusual behavior implies the presence of significant spin fluctuations of Bogoliubov quasiparticles, which are associated with possible nesting properties between BFSs.
\end{abstract}

\maketitle

\section*{Introduction}

The study of the iron-based superconductor, FeSe, has resulted in various topics, such as the appearance of superconductivity in the electronic nematic phase without magnetism, Bardeen-Cooper-Schrieffer Bose-Einstein-condensation (BCS-BEC) crossover, and Fulde-Ferrell-Larkin-Ovchinnikov (FFLO)  superconductivity at very high magnetic fields \cite{Matsuda}. FeSe has unique features among iron-based superconducting (SC) systems, such as  the formation of Fermi surfaces with unconnected small hole and electron pockets with very small Fermi energies, and the absence of magnetism in the nematic phase where four-fold symmetry is broken on the FeSe plane. These features are maintained by isovalent substitution in FeSe$_{1-x}$S$_x$ up to the nematic quantum critical point (QCP) ($x_{\rm c} \simeq 0.17$). In contrast to magnetism, the SC state spreads across a wide S-substitution regime \cite{Matsuura2017} and that developed in the nematic phase has a higher $T_{\rm c}$ ($ \sim $9 K) than that in the tetragonal phase ($x > x_{\rm c}$) with four-fold symmetry ($T_{\rm c}$$\sim$ 4 K). According to recent angle-resolved-photo-emission-spectroscopy (ARPES) measurements for pure FeSe, the hole pocket near the zone center ($ \Gamma$  point) exhibits an ellipsoidal shape, and its SC gap has a two-fold symmetric character with possible nodal points \cite{Hashimoto}. This implies that $s$- and $d$-wave components are significantly mixed in the nematic phase. With increasing S-concentration levels, the nematic fluctuations are strongly enhanced, and the transport properties exhibit non-Fermi liquid behavior.  Accompanied by such changes near the nematic QCP, a change in the SC-gap structure has been suggested based on the field dependence of the specific heat and thermal conductivity \cite{Sato2018}.  Moreover, a marked change in the SC gap has been observed in systematic scanning-tunneling-spectroscopy (STS) studies \cite{Hanaguri2018, Hanaguri}. The SC-gap spectrum showed a small zero-energy conductance below $x_{\rm c}$,  whereas it was significantly enhanced  above $x_{\rm c}$.

Recently, Setty, \emph{et al.} proposed a theoretical model for a system with spin-orbit coupling, multi-band structures, and time-reversal symmetry breaking (TRSB), and suggested that topologically protected nodal Fermi surfaces, referred to as Bogoliubov Fermi surfaces (BFSs), are induced in the SC state \cite{Setty, SettyPRB}. The appearance of the nonzero DOS in the specific heat and thermal conductance measurements in a heavily S-substituted regime can be explained by the formation of BFSs. TRSB required for BFSs was recently suggested from muon spin relaxation measurements \cite{Matsuura2023}. The existence of BFSs is a significantly important topic because it is deeply associated with SC pairing symmetry. Furthermore, whether interband interactions exist between BFSs remains an open question. We investigated the formation of BFSs in the SC state and interband interactions from the perspective of low-energy magnetic fluctuations using $^{77}$Se-nuclear magnetic resonance (NMR) down to 100 mK. Deep in the SC state, we found the presence of spin fluctuations of Bogoliubov quasiparticles, which are associated with possible nesting properties between BFSs.

\section*{Results and discussion}

Previous NMR measurements were performed mainly above $T_{\rm c}$ for pure bulk FeSe \cite{Baek2015, Wang, Wang2020, Wiecki2017a, Ishida, Zhou, Molatta, Vinograd, Chen} and S-substituted FeSe \cite{Wiecki2018, KuwaJPSJ, Rana, BaekQM, KuwaSR, Rana2}. Studies have primarily focused on the relationship between superconductivity, nematicity, and magnetic order above $T_{\rm c}$. We performed $^{77}$Se-NMR measurements in both the normal and SC states of FeSe$_{1-x}$S$_x$ ($0\leq x \leq 0.23$).  A single crystal was used for each S-substitution level, and a magnetic field of 6.0 T was applied parallel to the FeSe plane to exclude the formation of vortices in the plane. $T_{\rm c}$ was determined from AC susceptibility measurements utilizing an NMR tank circuit. The susceptibility data are presented in detail in Supplementary figure 1 (see Supplementary Note 1); $T_{\rm c}$ for $x \leq 0.12 $ was approximately 9 K, whereas $T_{\rm c}$ for $x$= 0.18 was 3 K at 6.0 T.  $^{77}$Se-NMR spectra for several S-substitution levels across $x_{\rm c} \sim 0.17$ are shown later.  Similarly to pure FeSe \cite{Baek2015}, the NMR spectra showed double peaks or edges below the nematic transition temperature, $T_{\rm nem}$, which merged into one peak with increasing S substitution.

\begin{figure*}[b]
\includegraphics[clip, width=0.8\columnwidth]{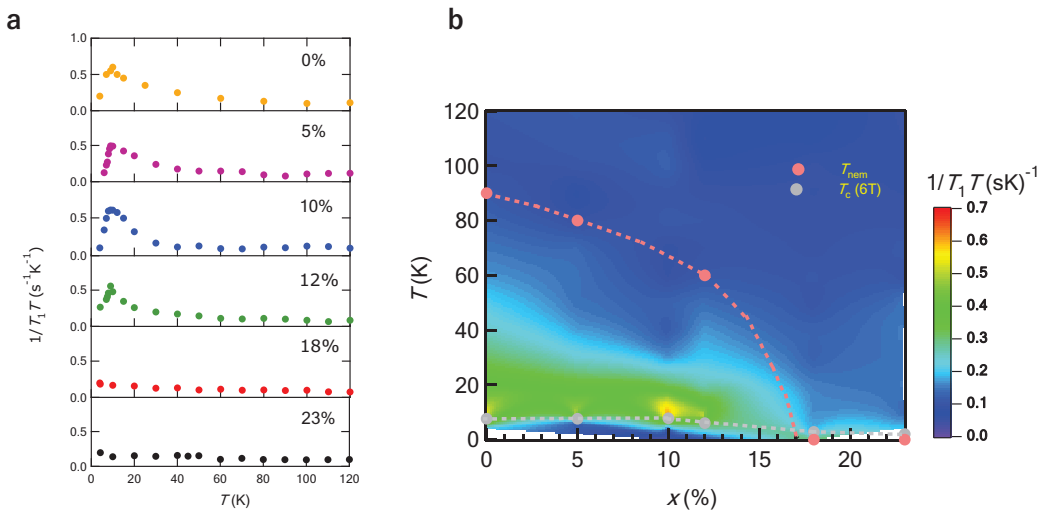}
\caption{\label{fig:epsart} {\bf $^{77}$Se relaxation rate ($1/T_1$) divided by temperature ($T$) of FeSe$_{1-x}$S$_x$ }  {\bf a,}   $T$ dependence of $1/T_1T$ at high temperatures. The magnetic field of 6.0 T was applied parallel to the FeSe plane. The data for 12\%-S substituted FeSe were already published \cite{KuwaJPSJ, KuwaSR}.  {\bf b,} Color plot of  $1/T_1T$. $T_{\rm nem}$ represents the nematic transition temperature, and $T_{\rm c}{\rm (6T)}$ represents the superconducting (SC) transition temperature measured by means of the AC susceptibility at 6.0 T.}
\end{figure*}

Figure 1a shows the evolution of the relaxation rate divided by temperature ($1/T_1T$) with respect to S substitution, and Fig. 1b is a color plot of Fig. 1a. $1/T_1T$ provides a measure of low-energy spin fluctuations.  In general, $1/T_1T$ is expressed as:

 \begin{equation}
	\frac{1}{T_{1}T}\sim \sum_{\textbf{q}} Im \chi(\textbf{q})
	\end{equation} where $\chi(\textbf{q})$ is the wave-number ($\textbf {q}$) dependent susceptibility. The $T$ dependence of $1/T_1T$ shown in Fig. 1a contrasts with that of the Knight shift which is discussed later. The Knight shift, which reflects the uniform susceptibility $\chi(0)$, monotonically decreases with decreasing temperature. The upturns of $1/T_1T$ toward $T_{\rm c}$ originate from spin fluctuations with $\textbf{q} \neq 0$.  As shown in Fig. 1b, spin fluctuations develop remarkably in the nematic phase and are strongly suppressed in the tetragonal phase. Spin fluctuations are associated with the topological configuration of electron and hole pockets and interband couplings. Two-dimensional Fermi surfaces theoretically obtained  in the tetragonal phase \cite{Yamakawa2017a} are schematically shown in Figs. 2b and 2c.  The enhancement of $\chi(\textbf{q})$ is expected at $\textbf{q} \sim (\pi, 0)$ in the nematic phase, as well as in the tetragonal phase, owing to the interband coupling between electron and hole pockets, namely, $\textbf{q}=(\pi, 0)$ nesting. However, the absence of a magnetic order implies that the nesting is not very strong. Such weak coupling is expected when orbital-selective coupling becomes important.  The experimental results in Fig. 1a are almost consistent with those observed by another study \cite{Wiecki2018, BaekQM}, wherein spin-fluctuation mediated pairing, that is, $s^{\pm}$-wave pairing was suggested because $T_{\rm c}$ reached a maximum for significantly strong spin fluctuations ($x \sim 0.1$) \cite{Wiecki2018}.

\begin{figure*}[b]
\includegraphics[clip, width=0.8\columnwidth]{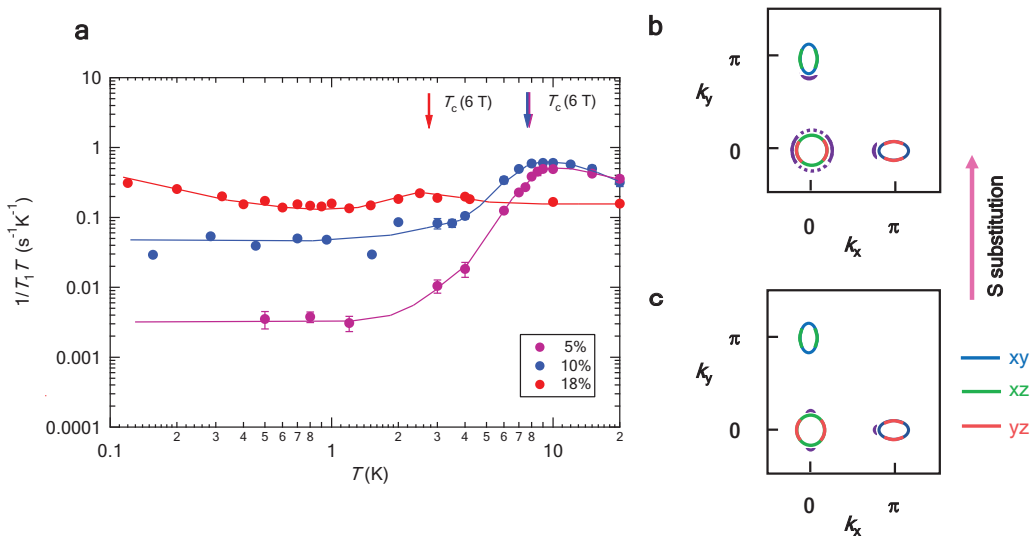}
\caption{\label{fig:epsart} {\bf $1/T_1T$ in the SC state and Bogoliubov Fermi surfaces (BFSs)} {\bf a,} Temperature dependence of $1/T_1T$ at low temperatures below $T_{\rm c}$ for several S-substitution levels. Arrows indicate $T_{\rm c}$ measured at 6.0 T. Error bars are estimated from those of $T_1$. All data points contain the error bars, although many of the error bars are smaller than the data points.  {\bf b,} Schematic diagram of two-dimensional Fermi surfaces obtained theoretically for the tetragonal phase together with contributions from three orbitals, $d_{xy}$, $d_{yz}$ and $d_{zx}$ colored in blue, red, and green, respectively \cite{Yamakawa2017a}. Sufficiently expanded BFSs expected below $T_{\rm c}$ are colored in purple \cite{Setty}. Recently, quasiparticle excitations with a finite gap has been observed for the directions shown by the dotted curves \cite{Okazaki}.  {\bf c,} BFSs expanded from the nodal points of pure FeSe are colored in purple \cite{Setty}. }
\end{figure*}

Figure 2a shows $1/T_1T$ at temperatures below $T_{\rm c}$ for several S-substitution levels crossing $x_{\rm c} \sim 0.17$, which is the main results in this study. The decrease in $1/T_1T$ below $T_{\rm c}$ is due to the opening of the SC gap. In conventional clean  superconductors, $1/T_1T$ should decrease to zero with decreasing temperature.  However, $1/T_1T$ for $x$=0.05 and 0.10 became constant at low temperatures. With further substitution  over $x_{\rm c}$,  $1/T_1T$ exhibited an upturn with decreasing temperature and the values became significantly larger than those for $x$=0.05 and 0.10.  Upturns of $1/T_1T$ observed below and above $T_{\rm c}$ are very rare in SC systems. The low $T$ behavior of $1/T_1T=constant$ suggests a residual DOS. As a cause of the residual DOS, (1) the impurity effect, (2) the Volovik effect, and (3) the coexistence of SC and normal states may be possible. First, the behavior of $1/T_1T = constant $  indicates an impurity effect in most cases. The impurity effect on $1/T_1$ has been studied theoretically at an early stage in the $s^{\pm}$-wave scenario \cite{BangYunkyu}.

For small impurity doping, the quasiparticle DOS is V-shaped  as a function of energy, and the $T$ dependence of $1/T_1T$ shows a $T^2$ dependence at temperatures below $T_{\rm c}$. For fairly large impurity doping, the quasiparticle DOS has a finite value at zero energy, and $1/T_1T$ exhibits a $T$-independent relation followed by a $T^2$ dependence with decreasing temperature. At first, the impurity effect appears to explain the low $T$ behavior of $1/T_1T$; however, this effect is excluded for the following reasons: (i) The upturn of $1/T_1T$ for $x$=0.18 is difficult to explain, and (ii) impurity doping levels should be remarkably different between $x$=0.05 and 0.10 because the constant value of $1/T_1T$ differs by approximately one order of magnitude. Therefore, $T_{\rm c}$ should be remarkably different between them. However, the value of $T_{\rm c}$ is almost the same between them. (iii) Furthermore, quantum oscillation measurements observed in a wide range of $x$ covering the tetragonal phase indicates the relatively clean nature of the samples \cite{ColdeaNPJ}. The Volovic effect can also induce an in-gap state similarly to the impurity effect. The DOS can be induced by an applied field owing to the Doppler shift of quasiparticle energy. The field dependence of $1/T_1T$ has been theoretically investigated for $s^{\pm}$- and $d$-wave cases \cite{Bangvolovik}. This effect should result in almost the same $1/T_1T$ values between x=0.05 and 0.10, because the strength of the applied field is the same and the SC gap or $T_{\rm c}$ is almost the same between them. Therefore, this effect is excluded as well as the impurity effect. Another possibility is the coexistence of SC and normal states. The coexistence in real space implies that SC and normal metallic domains coexist. However, such a case can be excluded by the rather homogeneous spectra observed in the STS measurements \cite{Hanaguri2018}. Instead of the coexistence in real space, the coexistence in momentum space such as the formation of  BFSs would be promising. BFSs that are expanded from the nodal points of pure FeSe and sufficiently expanded BFSs \cite{Setty} would play important roles for $x < x_{\rm c}$  and  $x > x_{\rm c}$, respectively, in view of the STS and specific-heat results \cite{Hanaguri2018, Sato2018}. The schematics of the former and latter are shown in Figs. 2c and 2b, respectively. In fact, BFSs should be much complicated because BFSs with two-fold symmetry have been suggested from recent ARPES measurments, although the measurements were performed for the tetragonal phase \cite{Okazaki}.

For small $x$, BFSs expanded from the nodal points are small, and the nesting is hardly expected, as shown in Fig. 2c. Therefore, the enhancement of $\chi(\textbf{q})$ or the upturn of $1/T_1T$ is hardly expected, despite that such enhancement or upturn is caused in the normal state due to the nesting between original Fermi surfaces. In such a case, flip flop of a nuclear spin due to scattering by quasiparticles becomes a major relaxation process, like the Korringa relation in conventional metals. The Korringa relation, $1/T_1T K_{\rm spin}^2 =constant$, where $K_{\rm spin}$ is the spin part of the Knight shift $K$, is applied for conventional metals with free electrons or weakly interacting electrons. $K$ is decomposed into $K_{\rm spin}$ and the orbital part $K_{\rm orb}$,

\begin{equation}
	K=K_{\rm spin}+K_{\rm orb},
    \end{equation}
where $K_{\rm spin}$ is proportional to the DOS and the uniform susceptibility $\chi$(0). $K_{\rm spin}$ and $K_{\rm orb}$ can be separated at high temperatures using the susceptibility data, $\chi$(0).  $K_{\rm orb}$ is $T$-independent and is approximately 0.26\% for $x$=0.12 \cite{KuwaSR}. For conventional superconductors, $K_{\rm spin}$ approaches zero as temperature is decreased owing to the SC gap. In case considered in this study, $K_{\rm spin}$ remains nonzero due to the DOS of BFSs. Therefore, $1/T_1T$ can be expressed using the Korringa relation as follows despite of the SC state:

  \begin{equation}
	\frac{1}{T_{1}TK_{\rm spin}^{2}}=\frac{4\pi k_{\rm B}}{\hbar}(\frac{\gamma_{\rm n}}{\gamma_{\rm e}})^{2}K(\alpha)
	\end{equation}
where $k_{\rm B}$  and $\hbar$ are the Boltzman and Planck constants, respectively, $\gamma_{\rm n (e)}$ is the gyromagnetic ratio of $^{77}$Se (electron), and $K(\alpha)$ is a function of the Stoner enhancement factor $\alpha$. The Stoner factor is expressed as $\alpha=I \chi$(0) where $I$ represents the interaction coupling. $K(\alpha)$ provides a measure of electron correlation. For ferromagnetic metals, $K(\alpha)<1$,  whereas for antiferromagnetic metals, $K(\alpha)>1$. Further, for free electrons, $\alpha$=0 and $K(\alpha)$=1. A detailed expression of $K(\alpha)$ is provided in Supplementary Note 2. As $1/T_1T$ for $x$=0.10 is one order larger than that for $x$=0.05, $K_{\rm spin}$ estimated using Eq. (2) is approximately three times larger than that for $x$=0.05. We estimated residual $K_{\rm spin}$ from the experimental results of $1/T_1T$, assuming that the interaction between quasiparticles is negligibly small, that is,  $K(\alpha)$=1.  The value of $K_{\rm spin}$ is estimated to be 0.015\%  for $x$=0.05. Subsequently, the ratio of $K_{\rm spin}$ between the SC and normal states is estimated to be 0.49, implying that the DOS of BFSs reaches almost half of the DOS observed in the normal state above $T_{\rm c}$.  In fact, certain quasiparticle correlations should exist ($K(\alpha)>1$), and the value mentioned above is overestimated.  To obtain a more precise estimation, a further theoretical analysis of $\chi(\textbf{q})$ is required.

\begin{figure*}[t]
\includegraphics[clip, width=0.8\columnwidth]{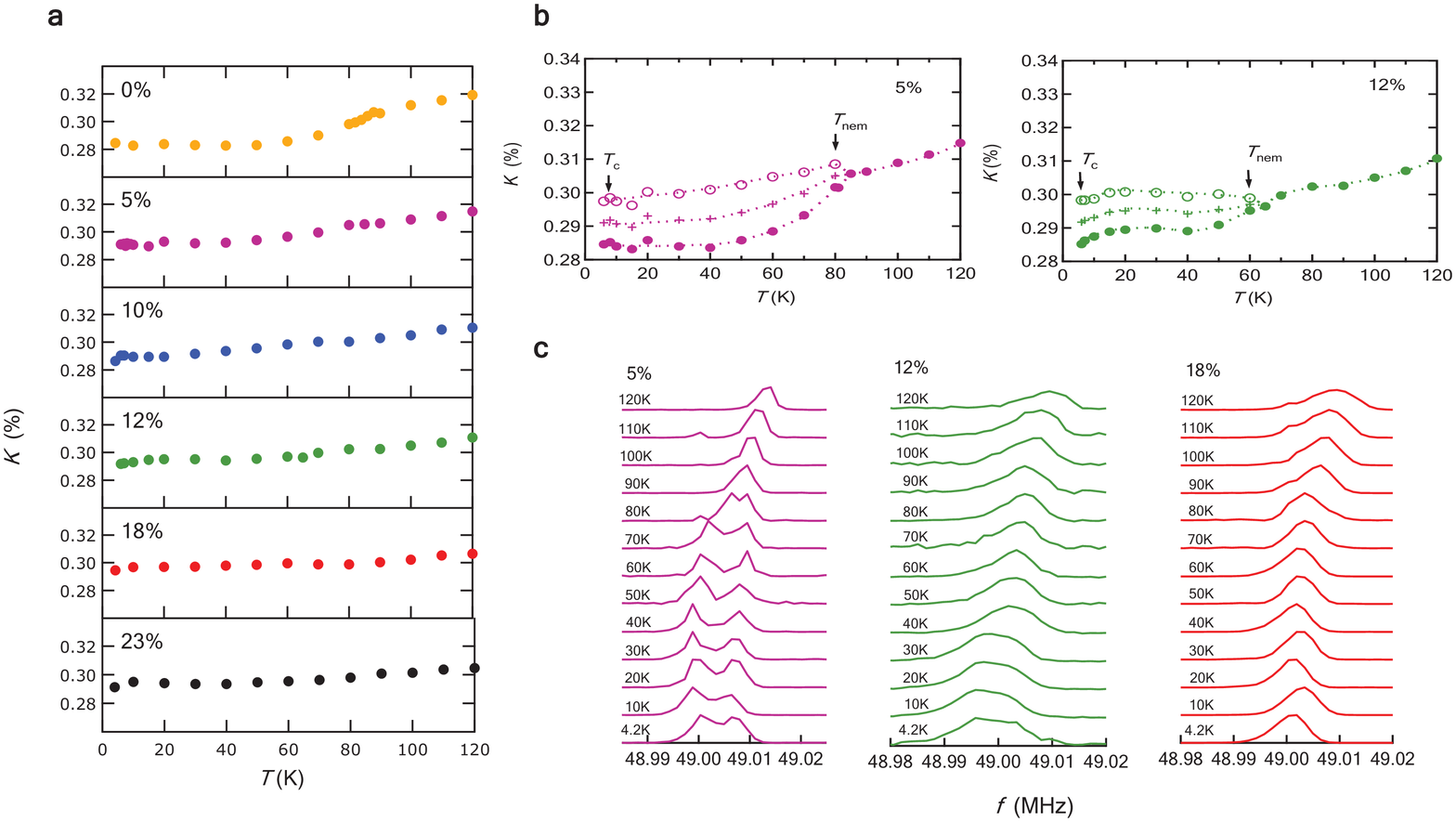}
\caption{\label{fig:epsart} {\bf $^{77}$Se Knight shift ($K$) of FeSe$_{1-x}$S$_x$ determined from nuclear magnetic resonance (NMR) spectra}  {\bf a,} Temperature dependence of $K$ at high temperatures.  The shifts were measured at the field (6.0 T) parallel to the FeSe plane.  In the nematic phase ($x < 0.17$), two peaks or edges appear as shown in Fig. 3c.  We plotted the average of two peaks for $K$ below $T_{\rm nem}$. The data for 12\%-S substituted sample ($x$=0.12) was cited from the previous works \cite{KuwaJPSJ, KuwaSR}. {\bf b,} $K$ below $T_{\rm nem}$ for $x$=0.05 and 0.12.  $T_{\rm nem}$ and $T_{\rm c}$ represent the nematic and superconducting transition temperatures, respectively. Two peaks in Fig. 3c and their average are shown in both panels. Both panels are expansions of Fig. 3a.  {\bf c,} $^{77}$Se-NMR spectra for $x$=0.05, 0.12, and 0.18. Two peaks appearing for $x$=0.05 and 0.12 merge into a single peak with increasing S-substitution level over $x_{\rm c}$ ($\geq$ 0.17).}
\end{figure*}

The estimation for $x$=0.05 and 0.10 is not applicable for $x$=0.18, because the Korringa relationship breaks. The nesting between sufficiently expanded BFSs can cause the enhancement of $\chi(\textbf{q})$. A drop just below $T_{\rm c}$ originates from the opening of the SC gap and  a rise just above $T_{\rm c}$ originates from spin fluctuations which are strongly suppressed in the tetragonal phase. Interestingly, the $T$ dependence of $1/T_1T$ below $T_{\rm c}$ is similar to that above $T_{\rm c}$, although the magnitude of the upturn becomes one third across $T_{\rm c}$. The $T$ dependence of $\chi(\textbf{q})$ at high temperatures originates from the nesting between original Fermi surfaces in the normal state. The reason why similar $\chi(\textbf{q})$ is reestablished below $T_{\rm c}$ is that sufficiently expanded BFSs reflect original Fermi surfaces and a similar nesting becomes possible.

We have estimated $K_{\rm spin}$ from $1/T_1T$, however, it seems straightforward to estimate from $K$. Figure 3a shows the evolution of $K$ with respect to S substitution. The results are almost consistent with those observed by another study \cite{Wiecki2018}.  In the nematic phase, two peaks appear reflecting two domains (the spectra of $x$=0.05 and 0.12 in Fig. 3c). In Fig. 3a, only the average of the two peaks is plotted for temperatures below $T_{\rm nem}$. The detailed shifts of each peak and their average are shown in Fig. 3b. Figure 3a certainly includes the information of $\chi(0)$ or $K_{\rm spin}$. These quantities can be obtained from $K-\chi$ plots at high temperatures. This procedure is possible because both quantities are $T$-dependent at high temperatures (Fig. 3a). However, at very low temperatures below $T_{\rm c}$, $K_{\rm spin}$ originating from the residual DOS and $K_{\rm orb}$ can not be separated from each other because both quantities are $T$-independent. Thus, a theoretical investigation is needed to separate $K_{\rm spin}$ from $K_{\rm orb}$. Even if $K_{\rm orb}$ is theoretically calculated, small residual $K_{\rm spin}$ $(< 0.03)$ is estimated from the raw data, $K (\sim 0.28)$ by subtracting large $K_{\rm orb}$ $(\sim 0.25)$. This procedure can potentially contain a significant error. Furthermore, $K_{\rm orb}$ may change below and above $T_{\rm nem}$ in the lightly S-substituted regime, which renders the estimation of residual $K_{\rm spin}$ much more challenging. The estimation of $K_{\rm spin}$ from $1/T_1T$ is advantageous because the information on $K_{\rm orb}$ is not needed.

The low-$T$ spin fluctuations above $x_{\rm c}$ can be attributed to the interband couplings between BFSs. However, the question arises as to whether the disappearance of the nematic order or revival of four-fold symmetry on the FeSe planes is essential for the couplings between BFSs. To investigate this problem, 12\%-S substituted FeSe at approximately 1 GPa may provide a clue because the $T$ dependence of $1/T_1T$ is similar to that of 18\%-S substituted FeSe, that is, the suppression of spin fluctuations has been observed for 12\%-S substitution  at approximately 1 GPa \cite{KuwaJPSJ, KuwaSR}. Further investigations at very low temperatures and high pressures are required for S-substituted FeSe ($x < x_{\rm c}$).

\section*{Conclusion}
In conclusion, we have observed anomalous spin fluctuations deep in the SC state for S-substituted FeSe over $x > x_{\rm c}$ via $1/T_1T$ measurements to 100 mK, which suggests the presence of Bogoliubov quasiparticles. The upturn of $1/T_1T$ below $T_{\rm c}$ is similar to that above $T_{\rm c}$, suggesting that $\chi(\textbf{q})$ does not significantly change across $T_{\rm c}$. Therefore, the upturn can be attributed to the enhancement of $\chi(\textbf{q})$ due to the nesting between BFSs. This means that the expansion of BFSs is significantly large. The low-$T$ spin fluctuations above $x_{\rm c}$ contrast with those below $x_{\rm c}$ arising from weekly interacting quasiparticles.  The present NMR results highlight the novel SC state with BFSs, having excitations at zero energy.

\section*{Methods}

We performed $^{77}$Se-NMR measurements at 6.0 T using a single crystal for each S-substitution level. Typical size is approximately 1.0 mm $\times$ 1.0 mm $\times$ 0.5 mm. We applied a magnetic field parallel to the FeSe planes to suppress the decrease in $T_{\rm c}$ and generation of vortecies. We performed pulsed-NMR measurements using a conventional spectrometer. NMR spectra were obtained by the Fast Fourier Transform (FFT) of a spin-echo signal. To detect a echo signal, we used 4-cycle pulse sequence. The relaxation time ($T_1$) was measured by the saturation-recovery method, and the data points of the recovery curve were fitted by a single exponential function.

\bibliography{References}


\section*{Acknowledgements }

The authors would like to thank S. Nagasaki and T. Takahashi for their experimental support. The present work was supported by Grants-in-Aid for Scientific Research (KAKENHI Grant No. JP18H01181), JST SPRING, Grant Number JPMJSP2110, a grant from the Mitsubishi Foundation, and a grant from The Kyoto University Foundation. This work was partly supported by Grants-in-Aid for Scientific Research (KAKENHI Grant Nos. JP22H00105, JP18H05227, JP19H00648, and JP18K13492) and by Innovative Areas “Quantum Liquid Crystals” (No. JP19H05824) from the Japan Society for the Promotion of Science.

\section*{Author contributions statement}

N. Fujiwara designed the NMR experiments.  Z.-Y. Yu, K. Nakamura, and K. Inomata carried out the NMR measurements. K. Matsuura, Y. Mizukami, S. Kasahara, Y. Matsuda, and T. Shibauchi synthesized the samples and performed the chemical analysis of the samples.  X. -L. Shen, and T. Mikuri operated the dilution refrigerator to cool down to 100 mK under the instruction of Y. Uwatoko.


\section*{{\large Supplementary Information}}


\subsection {Supplementary Note 1: Determination of $T_{\rm c}$ from the AC susceptibility}

We determined $T_{\rm c}$ from the resonance frequency $f_{\rm r}$ of the tank circuit attached to the head of an NMR probe. The frequency $f_{\rm r}$ was measured using a commercially available network analyzer. It is related to the AC susceptibility $\chi$ as $f_{\rm r}=1/\sqrt{LC(1+4\pi\chi)}$, where $C$ and $L$ represent the capacitance of a variable capacitor and the inductance of a coil wound onto the sample, respectively. We measured at a magnetic field of 6.0 T at a frequency approximately 50 MHz, which roughly corresponds to the NMR frequency of a free $^{77}$Se. As shown in Fig. 1, $f_{\rm r}$ increases gradually with decreasing temperature from room temperature, as $L$ of the coil gradually decreases during the cooling process. A drastic increase in $f_{\rm r}$ occurs at $T_{\rm c}$ owing to the Meissner effect. We determined $T_{\rm c}$s from the crossing points of dashed lines as shown in the panel for 18 \%.  $T_{\rm c}$s determined at 6.0 T for 5\%, 10\%, and 18\% were 7.6, 7.8, and 2.8 $K$, respectively, whereas those at zero field were 8.8, 8.7, and 4.0 $K$, respectively.

\setcounter{figure}{0}
\renewcommand{\figurename}{Supplementary figure}

\begin{figure*}[h]
\includegraphics [clip, width=0.7\columnwidth] {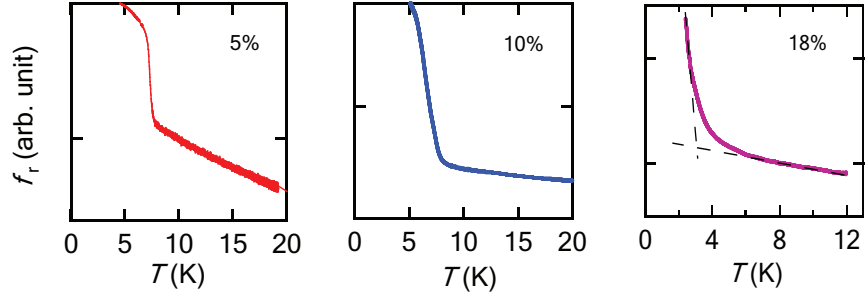}
\caption{\label{fig:wide} $T$ dependence of the resonance frequency $f_{\rm r}$ of the tank circuit for FeSe$_{1-x}$S$_x$ ($x$=0.05, 0.10, and 0.18) measured at 6.0 T. We used a single crystal for these measurements. }
\end{figure*}

\subsection {Supplementary Note 2: Detailed description of $K(\alpha)$}
The relaxation rate $1/T_1$ can be expressed using the dynamical susceptibility $\chi(\textbf{q}, \omega)$ based on the fluctuation-dissipation theory. For the hyperfine interaction, $H=A\textbf{\emph{I}}\cdot\textbf{\emph{S}}$, $1/T_1T$ is expressed as:

\begin{equation}
\frac{1}{T_{1}T}=\frac{2k_{\rm B}}{\hbar^{2}}(\frac{\gamma_{\rm n}}{\gamma_{\rm e}})^2\sum_{\textbf{q}}A_{\textbf{q}}A_{-\textbf{q}}\lim_{\omega\rightarrow 0}\frac{Im \chi(\textbf{q}, \omega)}{\omega} \tag{S1} \\
\end{equation} where $A_{\textbf{q}}=-<\textbf{k}+\textbf{q}|\frac{A}{\gamma_{n}\hbar}|\textbf{k}> $ and both $\textbf{k}$ and $\textbf{k}+\textbf{q}$ represent the wave numbers on the Fermi surfaces. $\gamma_{\rm n (e)}$ is the gyromagnetic ratio of $^{77}$Se (electron). For free electrons, $1/T_1T$ is expressed as
\begin{equation}
\frac{1}{T_{1}T}=\frac{\pi A^2}{\hbar N^2}N(\varepsilon_{\rm F})^2k_{\rm B} \tag{S2}
\end{equation} where $N(\varepsilon_{\rm F})$ represents the density of states at the Fermi level and N is the total number of sites. The spin part of the Knight shift $K_{\rm spin}$ is proportional to $N(\varepsilon_{\rm F})$ as well as the uniform susceptibility $\chi_0$ (=$\mu_{\rm B}^{2}N(\varepsilon_{\rm F})$) :
\begin{equation}
K_{\rm spin}=\frac{A\gamma_{\rm e}}{2N\gamma_{\rm n}}N(\varepsilon_{\rm F}). \tag{S3}
\end{equation} In the case that electrons interact with each other, the uniform susceptibility $\chi$ is expressed using $\chi_0$  as
\begin{equation}
\chi=\frac{\chi_0}{1-\alpha}  \tag{S4}
\end{equation} where $\alpha=I\chi_0$ and $I$ is the electron-electron interaction. The shift $K_{\rm spin}$ and $1/T_1T$ are expressed as
\begin{equation}
K_{\rm spin}=\frac{A\gamma_{\rm e}}{2N\gamma_{\rm n}}N(\varepsilon_{\rm F})\frac{1}{1-\alpha}  \tag{S5}
\end{equation}
\begin{equation}
\frac{1}{T_{1}T}=\frac{\pi A^2}{\hbar N^2}N(\varepsilon_{\rm F})^2k_{\rm B}<\frac{1}{(1-I\chi_0(\textbf{q}))^2}>_{\rm FS}  \tag{S6}
\end{equation} where $\chi_0({\emph{\textbf{q}}})$ represents the  $\textbf{q}$-dependent susceptibility for free electrons and $<...>_{\rm FS}$ represents the average over $\textbf{q}$ connecting two points on the Fermi surfaces. Using these equations, the Korringa relation can be extended as
\begin{equation}
\frac{1}{T_{1}TK_{\rm spin}^{2}}=\frac{4\pi k_{\rm B}}{\hbar}(\frac{\gamma_{\rm n}}{\gamma_{\rm e}})^{2}K(\alpha),    \tag{S7}
\end{equation}
\begin{equation}
K(\alpha)=(1-\alpha)^2<(\frac{1}{( 1-\alpha\chi_{0}(\textbf{q})/\chi_{0} )^2}>_{\rm FS}.   \tag{S8}
\end{equation} For free electrons, $K(\alpha)$=1 is obtained from $\alpha$=0.

\end{document}